\documentclass[12pt,notoc]{JHEP} 

\newcommand{\be}{\begin{equation}}
\newcommand{\ee}{\end{equation}}
\newcommand{\bea}{\begin{eqnarray}}
\newcommand{\eea}{\end{eqnarray}}
\newcommand{\bref}[1]{(\ref{#1})}
\newcommand{\tfrac}[2]{{\textstyle\frac{#1}{#2}}}
\newcommand{\cF}{\mathcal{F}}
\newcommand{\Lagb}{\mathcal{L}_\mathrm{b}}
\newcommand{\amax}{a_\mathrm{max}}
\newcommand{\rhob}{\rho_\mathrm{b}}

\newcommand{\pb}{p_\mathrm{b}}
\newcommand{\sgn}{\,\mathrm{sgn}\,}
\newcommand{\const}{\mathrm{const.}}

\title{\bf Cosmological Brane World Solutions with Bulk Scalar Fields}

\author{Stephen C. Davis\\ 
Department of Physics, University of Wales Swansea, \\ 
Singleton Park, Swansea, SA2 8PP, Wales \\ 
{E-mail:} \email{S.C.Davis@swansea.ac.uk} }

\abstract{
Cosmological brane world solutions are found for five-dimensional bulk
spacetimes with a scalar field. A supergravity inspired method for
obtaining static solutions is combined with a method for
finding brane cosmologies with constant bulk energies. This provides
a way to generate full (bulk and brane) cosmological solutions to
brane worlds with bulk scalar fields. Examples of these solutions, and
their corresponding Friedmann equations, are discussed. 
}


\preprint{SWAT/309}

\begin{document}

\section{Introduction}

There has been considerable interest in the idea that our
universe may be a 3-brane embedded in a higher dimensional
bulk space. This possibility has been motivated by ideas in
string/M-theory. All the usual standard model fields are confined
to the brane, while gravity can propagate in the bulk.
So far, gravity experiments have not detected any higher dimensional
effects, so if this `brane world' scenario is to be a model of our
universe, they must be suppressed.
In a scenario considered by Randall and Sundrum~\cite{R+S}, a positive
tension 3-brane is embedded in a five-dimensional bulk space with a
negative cosmological constant. If the brane tension and bulk energy
density are chosen appropriately, the standard four-dimensional gravity
will arise from graviton bound states on the brane. The
gravitons which are free to propagate in the bulk give corrections to this.

This idea was extended to an evolving universe by Bin\'etruy,
Deffayet, Ellwanger and Langlois~\cite{BWcos1,BWcos2}. As well as the usual
brane energy density there is also a brane cosmological constant. If
this is chosen appropriately it will cancel the bulk energy's
contribution on the brane and give a cosmology whose
late time Friedmann equation is approximately the same
as in the standard Friedmann-Robertson-Walker universe. Thus the
brane cosmology predictions for the current evolution of the universe
are as normal. However, the initial evolution of the universe is
dominated by the non-standard terms arising from brane effects.

In the Randall-Sundrum brane world scenario the bulk contains only a
cosmological constant. String/M-theory suggests that it will also contain
scalar fields, which are free to propagate through the bulk~\cite{BWact}. 
The evolution of these fields may have interesting cosmological
effects. For example, if the bulk energy density varies with time, the
effective brane cosmological constant will also be
time-dependent. This could give alternative mechanisms for inflation and
quintessence.

Solving the field equations for a brane world cosmology with scalar
fields is far more complicated than in the constant bulk case. One
approach is to express the five-dimensional Einstein equations in terms
of four-dimensional tensors on the brane~\cite{proj4}, and then use
boundary conditions to determine them~\cite{projcos}.  Unfortunately
this approach does not fully determine all the tensors. Progress can be
made by making assumptions about the bulk solution~\cite{projcos,
fluidbulk, anneSG}, but it is not obvious that such
assumptions are justified, or self-consistent. To determine which
solutions of the four-dimensional Einstein equations are allowed, the full
five-dimensional theory must be solved.

Full bulk solutions for static brane worlds have been found in some
special cases. If the scalar potential takes a supergravity-like form,
the field equations can be reduced to first order equations, which can
be solved with relative ease~\cite{BWsol1,BWsol2}. In this paper
I will combine that method with the work of 
Bin\'etruy {\em et al.}~\cite{BWcos2} to give cosmological solutions.

Conventions and background are outlined in section~\ref{sec bwcos}. 
In section~\ref{sec sol} I extend the results of 
ref.~\cite{BWcos2} to a special class of brane world cosmologies with
bulk scalar fields. In section~\ref{sec sg} the supergravity inspired
static solution generating technique is generalised to these brane
cosmologies. Some examples of this type
of solution are considered in section~\ref{sec exp}, and their bulk behaviour
is discussed in section~\ref{sec bulk}. The results are summarised in
section~\ref{sec conc}.

\section{Five-dimensional Brane Worlds}
\label{sec bwcos}

We will consider a theory with the effective five-dimensional action
\be
S=\int_\mathrm{bulk} d^5x \sqrt{-g}
\left\{\frac{1}{2\kappa^2}R - \frac{1}{2}(\nabla \phi)^2 - V(\phi) \right\}
- \int_\mathrm{brane} d^4x \sqrt{-h} \left\{ \frac{1}{\kappa^2}[K] 
+ \Lagb(\phi) \right\} \ .
\ee
$g_{AB}$ is the bulk metric, and $h_{AB}$ is the induced metric on the
brane. They are related by $h_{AB} = g_{AB} - N_A N_B$, where $N^A$ is the
unit vector normal to the 3-brane. The covariant derivatives
corresponding to $g_{AB}$ and $h_{AB}$ are $\nabla_{\! A}$ and $D_{\! A}$ 
respectively.
The Gibbons-Hawking boundary term, $[K]$, is required to give a
well-formulated action. $[K]$ is the change in $K = h^{AB}K_{AB}$ across 
the brane, and $K_{AB} = h^C_A h^D_B \nabla_{\! (C} N_{D)}$ is the extrinsic
curvature. $\kappa^2 = 8\pi/M_5^3$, where $M_5$ is fundamental
five-dimensional Planck mass.
$\Lagb(\phi)$ is the Lagrangian density for all the fields on the brane. It
includes the `ordinary' standard model
fields which are confined to the brane, and the brane
potential energy for the bulk scalar $\phi$.

Varying the action gives the bulk equations of motion 
\be
\nabla^2 \phi = \frac{d V}{d \phi} \ ,
\label{phibulk}
\ee
\be
G^A_B = \kappa^2 T^A_B = \kappa^2 \left( \nabla^A \phi \nabla_{\! B} \phi 
- \delta^A_B [\tfrac{1}{2}(\nabla\phi)^2 + V(\phi)] \right) \ .
\label{Tbulk}
\ee

The brane part of the action gives the jump conditions~\cite{BWact}
\be
\left[N^A  \nabla_{\! A} \phi\right] = 
\frac{\delta \Lagb}{\delta \phi} \ ,
\label{jump1}
\ee
\be
\left[K_{AB} - K h_{AB}\right] = -\kappa^2 t_{AB} \ ,
\label{jump2}
\ee
where $t_{AB}$ is the energy momentum tensor corresponding to $\Lagb(\phi)$.

The Codacci equation relates a projection of $G_{AB}$ to $K_{AB}$
\be
N^A G_{AB} h^B_C = D_{\! A} K^A_C - D_{\! C} K \ .
\label{codacci}
\ee
Combining this with \bref{Tbulk} and substituting it into the jump conditions
\bref{jump1} and \bref{jump2} implies 
\be
D_{\! A} t^A_C = - \frac{\delta \Lagb}{\delta \phi} D_{\! C} \phi \ .
\label{emcon1}
\ee
This is the effective energy-momentum conservation equation on the brane.

Following ref.~\cite{BWcos1} we consider a five-dimensional metric of the form
\be
ds^2 = g_{AB}dx^A dx^B = 
-n^2(t,y)dt^2 + a^2(t,y) \Omega_{ij} dx^i dx^j + b^2(t,y) dy^2
\label{metric}
\ee
where $\Omega_{ij}$ is a three-dimensional metric of space
with constant curvature $k=-1, 0, 1$. The coordinates along the brane
are denoted $x^\mu$ and $y=x^5$ is the
fifth bulk coordinate. The brane is located at $y=0$.
Differentiation with respect to $y$ and $t$ will be denoted by primes
and dots respectively. The brane's normal is
$N^A = \delta^A_5 b^{-1}$, and the extrinsic curvature is
$K^A_B = \mathrm{diag}(n'/(nb),\delta^i_j a'/(ab),0)$.

Assuming that the brane matter is describable by a perfect fluid, \\ 
$t^\mu_\nu = \mathrm{diag}(-\rhob,\pb,\pb,\pb)$. The jump conditions
\bref{jump1} and \bref{jump2} are then
\be
\frac{[\phi']}{b_0} = \frac{\delta \Lagb}{\delta \phi} 
\ ,
\label{phijmp}
\ee
\be
\frac{[a']}{a_0b_0} = -\frac{\kappa^2}{3} \rhob \ ,
\label{ajmp}
\ee
\be
\frac{[n']}{a_0b_0} = \frac{\kappa^2}{3}(3\pb + 2\rhob) \ ,
\label{njmp}
\ee
The subscript 0 denotes quantities at $y=0$, i.e.\ on the brane, and 
$[Q] = Q(y=0^+) - Q(y=0^-)$ is the change in $Q$ across the brane. If
we assume that the bulk solution is symmetric across the brane, i.e.\
invariant under $y \leftrightarrow -y$, then
$Q(0^-) = -Q(0^+)$ and (\ref{phijmp}--\ref{njmp}) can be
used to determine $\phi'$, $a'$ and $n'$ at $y=0$. 

The energy-momentum conservation equation \bref{emcon1} becomes
\be
\dot \rhob + 3 \frac{\dot a_0}{a_0}(\rhob +\pb)
= \frac{\delta \Lagb}{\delta \phi} \dot \phi_0 \ .
\label{emcon2}
\ee
The right-hand side of this equation gives the amount of energy
non-conservation of the matter fields on the brane.

\section{Brane Cosmology with Scalar Fields}
\label{sec sol}

Using the Gauss-Codacci formalism it is possible to express the
five-dimensional Einstein equations in terms of the
effective four-dimensional Einstein tensor, the extrinsic curvature,
and the electric part of the Weyl tensor~\cite{proj4}. Using the jump
conditions the extrinsic curvature is found in terms of the brane
energy-momentum. The Weyl tensor cannot be found in this way, so another
approach is needed to find full bulk solutions.
I will consider a special class of solutions for which the
five-dimensional field equations can be simplified and solved using
generalisations of the methods of refs.~\cite{BWcos2} and \cite{BWsol2}.

In order to simplify the Einstein equations, it is useful to introduce
the quantity
\be
\cF(t,y) = -\left(\frac{\dot a}{a n}\right)^{\!\! 2} 
+ \left(\frac{a'}{a b}\right)^{\!\! 2} \ .
\label{Fdef}
\ee
In the bulk the components of the Einstein tensor are related to $\cF$ by
\be
G^0{}_0 - \frac{\dot a}{a'} G^0{}_5 
	= \frac{3}{2a^3 a'}\partial_y(a^4\cF) - \frac{3k}{a^2}
\label{G00}
\ee
\be
G^5{}_5 - \frac{a'}{\dot a} G^5{}_0 
	= \frac{3}{2a^3 \dot a}\partial_t(a^4\cF) - \frac{3k}{a^2} \ .
\label{G55}
\ee
This extends a simplification used in ref.~\cite{BWcos2}. In that
paper the bulk energy was taken to be constant. The left hand sides of
\bref{G00} and \bref{G55} are then equal, and it is simple to show
that they imply $\cF=\cF(a)$. The Einstein equations can then be
solved with relative ease.

When scalar fields are introduced it is clear, after substituting in
\bref{Tbulk}, that the left hand sides of \bref{G00}
and \bref{G55} are not generally equal. However if we consider a
special class of solutions with $\phi=\phi(a)$, the two expressions
will be the same. Hence, as with the
constant bulk case, $\cF'/a' = \dot \cF/\dot a$, and so
$\cF=\cF(a)$ too. Both \bref{G00} and \bref{G55} then reduce to
\be
a\frac{d\cF}{da} + 4 \cF 
+ \frac{\kappa^2}{3}\cF\left(a \frac{d\phi}{da}\right)^{\!\! 2} 
+ 2 \frac{\kappa^2}{3} V(\phi) - \frac{2k}{a^2} = 0 \ .
\label{Feq1}
\ee
The other diagonal components of the Einstein equation 
($G_{ij} = \kappa^2 T_{ij}$) are automatically satisfied as a result
of the Bianchi identities.

The standard brane cosmology discussed in ref.~\cite{BWcos2} is a
simple example of this type of solution. It has $\phi = \const$, 
$V = \const < 0$ and  
\be
\cF = -\frac{\kappa^2}{6} V + \frac{k}{a^2} - \frac{\mathcal{C}}{a^4} \ .
\label{constbulk}
\ee
where $\mathcal{C}$ is an arbitrary integration constant.

On the brane the effective four-dimensional Friedmann equation can be
determined from $\cF$. Assuming the $Z_2$ symmetry, \bref{Fdef} and
\bref{ajmp} imply
\be
\frac{\dot a_0^2}{a_0^2} = \frac{\kappa^4}{36}\rhob^2 - \cF_0 \ .
\label{Fried1}
\ee
The time has been scaled so that $n_0=1$. $t$ is thus the proper
cosmological time on the brane.

When $\phi = \phi(a)$, the field equation \bref{phibulk} reduces to
\be
\nabla^2 \phi = a^2\cF
 \left(\frac{d^2\phi}{da^2} + \frac{1}{a}\frac{d\phi}{da}\right)
+ \left[(G^0{}_0+G^5{}_5)\frac{a}{3} + \frac{2k}{a}\right] \frac{d\phi}{da}
= \frac{dV}{d\phi} \ .
\label{phi}
\ee

Substituting $G_{AB}=\kappa^2T_{AB}$ and \bref{Feq1} into
\bref{phi} gives
\be
\cF \left(a\frac{d}{da}\right)^{\!\! 2} \phi + 
\left[a \frac{d\cF}{da} + 4\cF
+ \frac{\kappa^2}{3}\cF\left(a\frac{d\phi}{da}\right)^{\!\! 2}\right] 
a\frac{d\phi}{da} - \frac{d V}{d \phi} = 0 \ .
\label{Feq2}
\ee
Thus the original partial differential field equations have been
reduced to two ordinary differential equations.

\section{Supergravity-style Solutions}
\label{sec sg}

For certain choices of $V$ it is possible to generate solutions to the
field equations using simpler first order equations. 
If $k=0$ and $V$ is a supergravity-style potential
\be
V(\phi) = \frac{1}{8} \left(\frac{dW}{d\phi}\right)^{\!\! 2} 
- \frac{\kappa^2}{6}W^2 \ , 
\label{SG1}
\ee
then the field equations \bref{Feq1} and \bref{Feq2} are satisfied if
\be
\cF = \frac{\kappa^4}{36}W^2
\label{SG2}
\ee
\be
a \frac{d\phi}{da} = - \frac{3}{\kappa^2 W} \frac{dW}{d\phi} \ .
\label{SG3}
\ee
These equations are a generalisation of those discussed in
refs.~\cite{BWsol1} and \cite{ BWsol2}, where static ($t$ independent)
brane worlds were considered. The corresponding first order equations
discussed in those papers can be obtained by setting $b=1$,
$n=a=e^{A(y)}$.

If we were dealing with a supergravity theory then $W(\phi)$ would be the
superpotential. However supergravity is not required for solutions of
(\ref{SG1}--\ref{SG3}) to satisfy \bref{Feq1} and \bref{Feq2}. All
that is needed is the above form of the potential. For convenience I
will refer to $W$ as the superpotential whether the underlying theory
includes supergravity or not.

The Friedmann equation \bref{Fried1} becomes
\be
\frac{\dot a_0^2}{a_0^2} = \frac{\kappa^4}{36}[\rhob^2 - W_0^2] \ .
\label{Fried2}
\ee
Using \bref{SG3} the rate of change of $\phi$ on the brane can also be found.
\be
\dot \phi_0^2 = \frac{1}{4}\left(\frac{dW_0}{d\phi}\right)^{\!\! 2}
\left[\frac{\rhob^2}{W_0^2} - 1\right] \ .
\label{dphidt}
\ee

The equation \bref{SG3} relates the two jump conditions \bref{phijmp}
and \bref{ajmp}. If the bulk is $Z_2$ symmetric, the jump conditions
are only consistent if
\be
\frac{\delta \Lagb}{\delta \phi} 
= \left(\frac{1}{W}\frac{dW}{d\phi}\right)_{\!\! 0} \rhob \ .
\label{dLag}
\ee
Thus the energy on the brane is not conserved in this type of brane cosmology.
In order for $\Lagb$ is to satisfy the above condition, there must be
$\phi$ dependent couplings in the standard model Lagrangian. Some of
the models discussed in ref.~\cite{projcos} (or generalisations
of them) may be suitable. 

Combining \bref{dLag} with \bref{emcon2} gives the new energy
conservation equation
\be
\dot \rhob + 3 \frac{\dot a_0}{a_0}(\rhob +\pb) 
= \frac{\dot W_0}{W_0} \rhob \ .
\label{rho2}
\ee
Thus if $\dot W_0/W_0$ is negative, energy will leak off the brane. If it
is positive, energy will be pushed out of the bulk and onto the brane.

In the usual brane cosmology, the energy density $\rhob$ consists of
a cosmological constant part and an ordinary matter part. The simplest
generalisation of this is
\be
\rhob = \lambda W_0 + W_0 \rho \ ,
\ee
where $\rho$ is proportional to the energy density of the ordinary
matter. It has the same evolution as in the standard cosmology, i.e.\
$\rho \propto a^{-3(1+w)}$ if $p=w\rho$. More complicated solutions of
\bref{rho2} will also exist, but I will not consider them here.

The effective Friedmann equation~\bref{Fried2} then becomes
\be
\frac{\dot a_0^2}{a_0^2} =\frac{\kappa^4}{36}W_0^2 
\left[(\lambda^2-1) + \rho^2 + 2 \lambda \rho \right]  \ ,
\label{Fried3}
\ee
It resembles the usual brane cosmology, but with a time dependent
gravitational coupling strength. If $W$ tends to a suitable constant
at large $t$ then a conventional brane world Friedmann equation will be
obtained. If $\lambda=1$ this will give the same late time evolution
as the standard cosmology, i.e.\ $\dot a_0^2/a_0^2 \propto \rho$. 

By considering the brane Friedmann equation~\bref{Fried3}, we can rule
out many choices of $W$. If $W$ does not tend to a constant, then the
solution's cosmological evolution will not generally be compatible with
observations. It is still possible that such solutions will agree with
the standard cosmology at some intermediate time. However, we expect
the typical time-scale for these solutions to be of order $M_5^{-1}$, so
without significant fine-tuning of the solution
the period of acceptable evolution will be too short. Even if the
Friedmann equation does have appropriate late-time behaviour, there is
no guarantee that the effective gravitational interactions of particles
on the brane will take the standard form. By considering the effective
four-dimensional gravity, we will be able to further constrain
$W$. However in this paper I will concentrate on the implications of the
non-standard Friedmann equation.

It is worth noting that (\ref{SG1}--\ref{SG3}) still satisfy equations
\bref{Feq1} and \bref{Feq2} if we change the signs of $\cF$ and $V$.
In this case \bref{Fried2} becomes
\be
\frac{\dot a_0^2}{a_0^2} = \frac{\kappa^4}{36} [W_0^2 +\rhob^2] \ .
\ee
This can also be rewritten as \bref{Fried3}, but with $(\lambda^2-1)$
replaced with $(\lambda^2+1)$. Thus these alternative solutions will not
give the correct late time behaviour, because of the large
effective cosmological constant term.

\section{Solutions for Exponential Superpotentials}
\label{sec exp}

As an illustration I will consider superpotentials which are the sum of two
exponentials. These are well motivated by string/M-theory. A
time-independent example of one of them is discussed in
ref.~\cite{BWsol1}. With the aid of the transformation $\phi \to \pm
\phi +\const$ they can all be reduced to 
\be
W = c\left[\frac{e^{-\alpha_1 \phi}}{\alpha_1}
+ s \frac{e^{\alpha_2 \phi}}{\alpha_2}\right] \ ,
\label{W12}
\ee
with $s = \pm 1$ and $\alpha_1 \geq |\alpha_2|$, or if $\alpha_2 =0$,
\be
W = c\left[\frac{e^{-\alpha_1 \phi}}{\alpha_1} + \frac{s}{\kappa} \right] \ .
\label{W1}
\ee
Since the above superpotentials are loosely inspired by string theory,
it is natural for the coefficients $\alpha_i$ to be of order $\kappa$, as
in ref.~\cite{BWsol1}.

If $s=+1$ and $\alpha_2 \neq 0$ then $W$ has a minimum at $\phi=0$. 
If $s\alpha_2 < 0$ then $W$ has a zero at 
$\phi=\phi_\star=(\alpha_1+\alpha_2)^{-1} \ln(-s\alpha_2/\alpha_1)$,
or $\phi_\star = - \alpha_1^{-1} \ln \alpha_1$ if $s=-1$ and $\alpha_2=0$.

The corresponding potentials obtained from \bref{SG1} are
\be
V = \frac{c^2}{8}\left[
\left(1-\frac{4\kappa^2}{3\alpha_1^2}\right) e^{-2\alpha_1 \phi}
+ \left(1-\frac{4\kappa^2}{3\alpha_2^2}\right) e^{2\alpha_2 \phi}
- 2s \left(1+\frac{4\kappa^2}{3\alpha_1 \alpha_2}\right) 
e^{(\alpha_2-\alpha_1)\phi}\right] \ ,
\ee
and (for $\alpha_2 =0$)
\be
V = \frac{c^2}{8}\left[\left(1-\frac{4\kappa^2}{3\alpha_1^2}\right) 
e^{-2\alpha_1 \phi} - 2s \frac{4\kappa}{3\alpha_1} e^{-\alpha_1\phi}
- \frac{4}{3}\right] \ .
\ee
If $V$ is to be bounded from below then only some values of the
parameters are allowed.

The solutions of \bref{SG3} when $W$ is given by \bref{W12} are
\be
\ln (a/a_*) = -\frac{\kappa^2}{3\alpha_1 \alpha_2}
\ln\left| e^{\alpha_1 \phi} - s e^{-\alpha_2 \phi} \right| \ ,
\label{phisol}
\ee
where $a_*$ is an arbitrary constant.

When $\alpha_2 =0$, we will use the superpotential \bref{W1}
instead. In this case \bref{SG3} is solved by
\be
\ln (a/a_*) = \frac{\kappa}{3\alpha_1} (\kappa \phi + s e^{\alpha_1 \phi}) \ .
\label{phisolsp}
\ee
The evolution of each of these solutions falls in to one of six cases,
one for each choice of $s$ and the sign of $\alpha_2$. Since we are
trying to produce a big bang style cosmology, we will be interested in
solutions which start at $a=0$. There are also some solutions with
$\phi = \const$ and $dW/d\phi =0$, but these are effectively the same
as the constant bulk energy case \bref{constbulk}, so I will not
consider them further.

\subsection{Superpotential with a minimum, $\alpha_2 > 0$, $s=+1$}
\label{ss min}

The expression \bref{phisol} gives two
solutions in this case. As $a$ goes from 0 to $\infty$,
$\phi$ rolls down from either $+\infty$ or $-\infty$ to the minimum of
the superpotential at $\phi = 0$. While it is rolling down,
$\dot W/W < 0$, and so \bref{rho2} implies that energy leaks off the
brane.

The approximate behaviour of $\phi$ and $\cF$ in terms of $a$ can be found
when $a$ is small or large by using series and asymptotic expansions.  
Initially, when $a \approx 0$,
\be
\phi \approx \sgn(\phi) \frac{3\alpha_i}{\kappa^2}\ln(a_*/a) \ ,
\label{phiinf}
\ee
\be 
\cF \approx \frac{\kappa^4 c^2}{36 \alpha_i^2} 
\left(\frac{a_*}{a}\right)^{\!\! 6\alpha_i^2/\kappa^2}  \ .
\label{Finf}
\ee
where the index $i$ is equal to 1 if $\phi<0$ and 2 if $\phi>0$.
Substituting these expressions into \bref{Fried3}, we find that 
$a \propto t^{1/(4+3\alpha_i^2/\kappa^2)}$. The initial expansion is
thus faster than in the usual brane cosmology.

Later, when $a$ is large,
\be
\phi \approx \frac{1}{\alpha_1+\alpha_2} 
\left(\frac{a_*}{a}\right)^{\!\! 3\alpha_1\alpha_2/\kappa^2} \ ,
\label{phi0}
\ee
\be
\cF \approx \frac{\kappa^4 c^2}{36\alpha_1^2\alpha_2^2} 
\left[(\alpha_1+\alpha_2)^2 + \alpha_1\alpha_2 
\left(\frac{a_*}{a}\right)^{\!\! 6\alpha_1\alpha_2/\kappa^2}\right] \ .
\label{F0}
\ee
Thus when $t$ (and hence $a$) is large, $W^2 \to \const$ and so the
standard cosmology will be obtained from \bref{Fried3}, with order
$\rho a_0^{6\alpha_1\alpha_2/\kappa^2}$ corrections to the Friedmann
equation. Since $\dot W \to 0$ at late time, the amount of energy
leakage from the brane also tends to zero, as is 
required to agree with the standard cosmology.

\subsection{Superpotential with a zero, $\alpha_2 >0$, $s=-1$}
\label{ss zero}

In this case $\phi$ rolls from either $+\infty$ to $-\infty$, or vice
versa. At both ends of the solution $a \to 0$. We see from
\bref{phisol} that $a$ is bounded.

Initially $\dot W/W$ is negative, so energy drains off the brane [see
\bref{rho2}]. This continues until $\phi=\phi_\star$, where $W=0$. At
this point there is nothing left on the brane. This is also the point
at which the scale factor $a$ reaches its maximum value
($\amax$). After this the universe re-collapses. $\dot W/W$ is now
positive, so energy is sucked onto the brane. This continues until
$|\phi|$ and $|W|$ reach infinity, at which point $a$ has returned to 0. 

The approximate behaviours of $\phi$ and $\cF$ near $a=0$ are
given by \bref{phiinf} and \bref{Finf}. Near $a=\amax$
\be
\ln(a/\amax) \approx -\frac{\kappa^2}{6}(\phi-\phi_\star)^2 \ ,
\ee
\be
\cF \approx  \frac{\kappa^2 c^2 e^{-2\alpha_1\phi_\star}}{6 \alpha_1^2}
(\alpha_1+\alpha_2)^2 \ln(\amax/a) \ .
\label{Fmax}
\ee
By substituting \bref{Finf} and \bref{Fmax} into \bref{Fried3}, a
finite upper bound can found for the time taken to reach $\amax$.
While $\dot a=0$ there, \bref{dphidt} shows that 
$\dot \phi \neq 0$ and so $\phi$ continues to roll down (or up) the
superpotential. Thus the universe expands to its maximum size and then
re-collapses, all in finite time. For $\alpha_i \sim \kappa$, the
time taken for the universe to re-collapse will be of order $M_5^{-1}$. This
very short lifetime for the universe, and the way in which all energy (and
hence matter) flows off the brane, mean that this model is completely
incompatible with the standard cosmology.

The re-collapse of the universe is a common feature of solutions
which pass through zeros of $W$. If $W$ changes sign and 
$dW/d\phi \neq 0$, then \bref{SG3} implies that $d\phi_0/da_0$ changes sign at
this point. Equation \bref{dphidt} ensures that
$\dot \phi_0$ does not, provided $\rhob^2 > W_0^2$. Thus
$\dot a_0$ must also change sign. This is consistent with the
expression for $\dot a_0$, \bref{Fried2}. Once
$a_0$ starts to decrease, \bref{Fried2} implies that it will continue
to do so.

\subsection{Positive monotonic superpotential, $\alpha_2 < 0$, $s=-1$}
\label{ss mono}

There is just one solution for these parameters. The scalar field
$\phi$ starts at $-\infty$ when $a=0$, and then rolls down to
$+\infty$ as $a \to \infty$. The approximate large and small $a$
behaviour of $\phi$ and $\cF$ are given by \bref{phiinf} and
\bref{Finf}. While the initial behaviour of this solution is similar
to that of section~\ref{ss min}, $\cF \to 0$ at late time. Thus
the effective gravitational coupling tends to zero as $t \to \infty$.

This type of superpotential has been suggested as a possible candidate
for \\ quintessence~\cite{anneSG}. If we set $\rho=0$ and $\lambda^2 \neq 1$ in
\bref{Fried3}, we get a solution with a cosmological
constant which decreases with time. Unfortunately when the matter
fields are re-introduced, the solution is incompatible with the
standard cosmology, due to the asymptotically vanishing gravitational
coupling. Thus, this type of superpotential is not suitable after all,
at least not for the cosmological generalisation of supergravity
solutions discussed in this paper. 

\subsection{Superpotential with minimum and zero, $\alpha_2 < 0$, $s=+1$}
\label{ss m+z}

For these parameters $a=0$ when $\phi =0$ and $\phi=-\infty$. This gives
a total of three solutions. $\phi$ can roll from $-\infty$ to 0, or vice
versa. In both cases $a \to 0$ at the start and end of the
solutions. These are similar to the solutions of section~\ref{ss
zero}, in that the universe expands to a maximum value and then
re-collapses. As before the maximum value is reached when $W=0$.

The other solution, with $\phi >0$, starts at $\phi=0$ and
$a=0$. As $a$ increases to $\infty$, $\phi$ tends to $+\infty$.
The approximate behaviour of $\phi$ and $\cF$ there is given by
\bref{phi0} and \bref{F0} near $\phi=0$, and
is asymptotically given by \bref{phiinf} and \bref{Finf} when $|\phi|$
is large. Thus the $\phi > 0$ solution has $W^2 \to 0$ at large $t$
(and $a$), which is not cosmologically acceptable.

\subsection{Special cases $\alpha_2 = 0$, $s=\pm 1$}
\label{ss sp}

Asymptotic and series expansions of \bref{phisolsp} give the
approximate forms of $\phi$ and $\cF$ for large and small $a$ 
\be
\phi \approx \left\{ \begin{array}{cc} \displaystyle
\frac{3\alpha_1}{\kappa^2}\ln(a/a_*) & \phi < 0 \\ & \\ \displaystyle
\frac{1}{\alpha_1}\ln\left[\frac{3\alpha_1s}{\kappa}\ln(a/a_*) \right]
& \phi > 0 \end{array} \right. \ \ ,
\label{phiinf0}
\ee
\be 
\cF \approx \frac{\kappa^2 c^2}{36} \left\{ \begin{array}{cc} \displaystyle
\left[\frac{\kappa}{\alpha_1}
\left(\frac{a_*}{a}\right)^{\!\! 3\alpha_1^2/\kappa^2} + s\right]^2 
& \phi < 0 \\ & \\ \displaystyle
\left[\frac{\kappa^2}{3\alpha_1^2 \ln(a/a_*)} + 1\right]^2 & \phi > 0 
\end{array} \right. \ \ .
\label{Finf0}
\ee
Both these cases have $\phi$ rolling down the superpotential from
$-\infty$ to $+\infty$. If $s=-1$ the system will pass through a zero
of the superpotential, and as with the solution in section~\ref{ss
zero}, $a$ reaches its maximum value there, and then starts to
decrease.

The $s=+1$ case resembles the solution in section~\ref{ss min}, in
that $\cF \to \const$ at late $t$ (large $a$). Thus this type of
solution also provides a potentially viable cosmological model. The
initial expansion of the universe is 
$a \propto t^{1/(4 + 3\alpha_1^2/\kappa^2)}$. The late time Friedmann
equation resembles that of the usual brane cosmology with order 
$\rho/\ln a$ corrections.

\subsection{Summary}

The solutions considered in this section can be divided into two
distinct categories. Firstly there are the solutions which pass through a
zero of the superpotential. In this case the universe expands until
$W=0$. Up until this point energy is lost from the brane into the
bulk. By the time $W=0$ there is nothing left on the brane. After this
the universe contracts, and energy flows back onto the brane. This
continues until the universe re-collapses to $a=0$. This
scenario appears to have little resemblance to the standard cosmology. The
solutions described in sec.~\ref{ss zero}, sec.~\ref{ss m+z} (first two
cases), and sec.~\ref{ss sp} ($s=-1$) fall into this category.

Of course it is still possible that these solutions agree with the
standard cosmology at some intermediate period of their
evolution. However, unless the parameters are fine-tuned so that
$\alpha_i/\kappa$ are extremely small, this period will be far to short
to agree with observation.

From a cosmological point of view, the other solutions are more
promising. In this case $\phi$ rolls down to a minimum of $|W|$,
without passing through $W=0$. As the minimum is approached, 
$a \to \infty$. Energy also leaks off the brane in this case, but the
rate approaches zero as $a \to \infty$. 
If the final value of $W$ is finite, then the standard
cosmology can be obtained late in the evolution. The solutions in
sec.~\ref{ss min} and sec.~\ref{ss sp} for $s=+1$ fall into this
category, thus the only viable cosmological solutions with a
superpotential of the form \bref{W12} or \bref{W1} are those with
$s=+1$ and $\alpha_2 \geq 0$. The solutions of sec.~\ref{ss mono} and
sec.~\ref{ss m+z} (third case) also have $\phi$ rolling down to a
minimum, but since it has $W=0$ there, the energy remaining on the
brane tends to zero as $a \to \infty$.

It should be noted that although the solutions with a minimum of $|W|$
give the correct Friedmann equation (and hence agree with many
observations), their other properties could still conflict with
experimental results, and so they may not be acceptable after all. For
example, they may give the wrong effective gravitational forces for
objects on the brane.

\section{Bulk solutions}
\label{sec bulk}

So far I have not determined the $y$ dependence of the metric (or
$\phi$). Following ref.~\cite{BWcos2}, this can be found by solving
the non-zero off-diagonal component of the Einstein equations. When
$\phi = \phi(a)$ this is 
\be
G_{05} = 3\left(\frac{n' \dot a}{na} + \frac{\dot b a'}{ba} -
\frac{\dot a'}{a} \right) =\kappa^2 T_{05} 
= \kappa^2a' \dot a \left(\frac{d\phi}{da}\right)^{\!\! 2} \ .
\label{G05}
\ee
We can assume that $b=1$, in which case \bref{G05} is solved by 
\be
\frac{\dot a}{n} = \beta(t) \frac{\kappa^2}{6}
\exp\left\{ -\frac{\kappa^2}{3}\int
a\left(\frac{d\phi}{da}\right)^{\!\! 2} \, da\right\}
\ee
where the function $\beta$ is independent of $y$. If $V$ has a
supergravity-like form \bref{SG1}, the above expression simplifies to
\be
\frac{\dot a}{n} = \beta(t) \frac{\kappa^2}{6} W(\phi) \ .
\label{neq}
\ee
Since we want $n_0=1$, \bref{Fried2} implies that 
$\beta = a_0\sqrt{\rhob^2/W_0^2 - 1}$.

Substituting \bref{neq} into the definition of $\cF$ \bref{Fdef}
\be
(a')^2 = \frac{\kappa^4}{36} (\beta^2 + a^2) W^2(\phi) \ .
\label{abulk}
\ee
$a$ must also satisfy $a=a_0$ at $y=0$ and the jump condition
\bref{ajmp}. Since we are interested in a $Z_2$ symmetric solution,
$a$ will be a function of $|y|$.

The alternative solutions discussed at the end of section~\ref{sec sg} satisfy
\be
(a')^2 = \frac{\kappa^4}{36} (\beta^2 - a^2) W^2(\phi)
\ee
instead of \bref{abulk}. Solutions of this equation will have 
$a$ bounded in the bulk. Since they do not have the correct late time
cosmological evolution, I will not consider them further.

The corresponding time-independent bulk solutions can be obtained by
setting $\rhob=W_0$, so that $\beta=0$. Finding analytic solutions of
\bref{abulk} is not generally possible, although
solutions for a few special cases can be found.

If we take $\alpha_1=\alpha_2=\kappa/\sqrt{3}$ then the superpotential
used in section~\ref{ss min} simplifies to
\be
W=\frac{2c}{\alpha_1} \cosh(\alpha_1\phi) = \frac{c\sqrt{3}}{\kappa}
\left[\left(\frac{a_*}{a}\right)^{\!\! 2} + 4\right]^{1/2} \ .
\ee
The bulk behaviour of $a$ is
\be
a^2 = a_0^2\left[1 - \frac{\kappa^2\rhob}{3\mu}\sinh(\mu|y|)
+ \left(\frac{\rhob^2}{2W_0^2}+\frac{\kappa^4 W_0^2}{18\mu^2}\right)
(\cosh(\mu y)-1) \right] \ ,
\label{abulksol}
\ee
where $\mu=2c\kappa/\sqrt{3}$. 
$a^2$ has zeros if $\rhob  > W_0 > 3\mu/\kappa^2$, which is true for
cosmological solutions. The position of these singularities varies
with time. Unfortunately $\phi'$, $\dot \phi$ and $W$ are all
singular at these points, and so $T_{AB}$ is too. Thus the
singularities at $a=0$ are
naked curvature singularities. This problem could be resolved if
we have a compact bulk, and one or more other branes at suitable
distances from our brane~\cite{BWcos1,annesing}. This separation of the branes
will then be time dependent. In order for this scenario to work, the
other brane(s) must have negative energy densities, which may lead to
instabilities.

The bulk behaviour for the model discussed in section~\ref{ss zero}
can also be found when $\alpha_1=\alpha_2=\kappa/\sqrt{3}$.
This time the bulk solution of $a^2$ is given by \bref{abulksol}, but with
$\mu$ replaced by $i\mu$. As before there will be bulk singularities.

Finally, for $\alpha_1=-\alpha_2=\kappa/\sqrt{3}$, the bulk behaviour
of the solution in section~\ref{ss mono} is
\be
a^2 = a_0^2\left(1 - \frac{\kappa^2}{3} \rhob |y|
+ \frac{\kappa^4}{36} W_0^2 y^2\right) \ .
\ee
Again $a^2$ has zeros in the bulk when $\rhob > W_0$, and since
$T_{AB}$ is also divergent there, they are curvature singularities. In
this case $W \propto 1/a$ so the singularities occur when 
$|W| \to \infty$. This is also the case with the other two examples
above. This suggests that solutions which have $W$ bounded may be
curvature singularity free.

\section{Conclusions}
\label{sec conc}

By considering solutions of the special form $\phi = \phi(a)$ it is
possible to significantly simplify the field equations for a brane
world cosmology with a scalar field. Furthermore, if the theory has a
supergravity style potential, the simplified equations can be further
reduced to three first order differential equations. This gives a
generalisation of time-independent supergravity brane worlds to brane
cosmologies. For simplicity I have considered solutions with $Z_2$
symmetric bulks. 

Investigation of examples of these solutions suggests that the
evolution of these brane cosmologies starts and ends at points with
either $dW/d\phi = 0$ or $W=\pm\infty$ ($W$ being the
superpotential). If the solution passes through a point with $W=0$,
the universe will stop expanding, and re-collapse. The only solutions
which avoid this re-collapse are those where $\phi$ rolls down to a
minimum of $|W|$.

In order for these $Z_2$ symmetric, supergravity solutions to be
self-consistent, the matter fields on the brane must couple to the
bulk scalar in such a way that energy is not conserved on the
brane. In the case of solutions which
approach a minimum of $|W|$, the amount of leakage tends to zero, and
the standard cosmology is obtained at late times. With the other
solutions, all the energy will have leaked off by the time $W=0$ is
reached. After this energy flows back onto the brane until it re-collapses.
  
The bulk behaviour of the metric (and $\phi$) can also be
determined. Unfortunately for the exponential superpotentials
considered in this paper, they all have naked curvature singularities
in the bulk. The problems occur at points where $W=\pm\infty$. This
suggests that a solution which only passes through finite values of
$W$ may not have naked bulk singularities. The singularity problem
could also be resolved by adding additional branes to the bulk. Like
the singularities, the position of the branes would vary with time.

\section*{Acknowledgements}

I wish to thank W. B. Perkins for useful discussions, and PPARC for
financial support.

\end{document}